\documentclass{article}
\usepackage[utf8]{inputenc}
\usepackage{authblk}
\usepackage{setspace}
\usepackage[margin=1.25in]{geometry}
\usepackage{graphicx}
\graphicspath{ {./figures/} }
\usepackage{subcaption}
\usepackage{amsmath}
\usepackage{amssymb}
\usepackage{newunicodechar}
\newunicodechar{−}{\ensuremath{-}}

\usepackage[utf8]{inputenc}
\usepackage{amsmath}
\usepackage{amssymb}

\usepackage[
backend=biber,
style=numeric-comp,
sorting=none
]{biblatex}
\title{Ultrafast Magneto-Pressure Spectroscopy and Control of Correlated Phases 
	in a Trilayer Nickelate}
\author[a,b$\dag$]{Zhi Xiang Chong}
\author[a$\dag$]{Joong-Mok Park}
\author[a$\dag$]{Shuyuan Huyan}
\author[a,b]{Avinash Khatri}
\author[a]{Martin Mootz}
\author[c,1]{Xinglong Chen}
\author[c]{Daniel P. Phelan}
\author[a]{Liang Luo}
\author[d]{Ilias E. Perakis}
\author[c]{J. F. Mitchell}
\author[a,b]{Sergey L. Bud'ko}
\author[a,b]{Paul C. Canfield}
\author[a,b*]{Jigang Wang}

\affil[a]{Ames National Laboratory, U.S. Department of Energy, Ames, IA 50011, USA.}
\affil[b]{Department of Physics and Astronomy, Iowa State University, Ames, IA 50011, USA.}
\affil[c]{Materials Science Division, Argonne National Laboratory, Lemont, IL 60439, USA.}
\affil[d]{Department of Physics, University of Alabama at Birmingham, Birmingham, AL 35294, USA.}
\affil[1]{Present address: School of Physics, Southeast University, Nanjing, 211189, China.}
\affil[*]{Correspondence: jgwang@amelsab.gov; jgwang@iastate.edu}
\affil[$\dag$]{These authors contributed equally to this work.}

\date{}

\begin{document}
	
	\maketitle
	
	%%%%%% Abstract %%%%%%
	\begin{abstract}
		Ultrafast spectroscopy under simultaneous high pressure and magnetic field provides a versatile approach for investigating pressure-driven electronic instabilities and correlated phases, and for probing potential bulk superconducting behavior under extreme conditions. However, such an experimental platform has yet to be implemented, standing as a roadblock to a fuller understanding of nonequilibrium superconductivity and vortex-controlled quasi-particle (QP) dynamics.
		Here, we bridge this capability gap by developing high pressure (up to 40 GPa), high magnetic field (up to 7 T), cryogenic (down to 5 K) femtosecond spectroscopy, and using it to probe magneto-pressure evolution of quasiparticle dynamics in the trilayer nickelate $\mathrm{Pr}_4\mathrm{Ni}_3\mathrm{O}_{10}$. 
		We observe pronounced critical slowing down of QP relaxation at the charge-density-wave transition, which collapses under applied pressure. At higher pressures, the relaxation instead lengthens at low temperature, consistent with incipient superconducting correlations. 
		However, the negligibel magnetic-field-dependence up to 7~T and absence of vortex-induced pre-bottleneck dynamics--robust signatures observed in our controlled bulk superconducting samples--indicates that any superconducting state under the present pressure conditions is likely non-bulk, filamentary, or strongly inhomogeneous.  
		The magneto-pressure ultrafast capability opens a new avenue for resolving outstanding questions surrounding pressure-induced superconductivity and intertwined orders in correlated quantum materials.
		%Our results demonstrate the power of ultrafast spectroscopy under simultaneous high pressure and magnetic field, opening a new pathway to disentangle intertwined electronic orders in quantum materials.
		%for exploring the complex landscape of correlated materials with competing ground states.
		%These results highlight the power of combining ultrafast, high-pressure, and high-magnetic-field control to disentangle intertwined electronic orders. 
		
		%The abstract should be a single paragraph written in plain language that a general reader can understand. Do not include citations, figures, tables, or undefined abbreviations in the abstract. Any abbreviations that appear in the title should be defined in the abstract. The length should be 200 words and not exceed 250 words, to include: 
		%\begin{itemize}
		%    \item An opening sentence that states the question/problem addressed by the research AND
		%    \item Enough background content to give context to the study AND
		%    \item A brief statement of primary results AND
		%    \item A short concluding sentence.
		%\end{itemize} 
	\end{abstract}
	
	
	%%%%%% Main Text %%%%%%
	
	\section{Introduction}
	The discovery of high-temperature superconductivity in layered nickelates~\cite{14,15,16,17,18,19,20,21,22,23,24,25,26} has sparked a new wave of exploration into unconventional superconductors, following the earlier breakthroughs in cuprates~\cite{27,28} and iron-based superconductors~\cite{29,30,31,56}. Among those, the double- and triple-layered Ruddlesden–Popper (RP) phases--especially $\mathrm{La}_3\mathrm{Ni}_2\mathrm{O}_{7}$ and $\mathrm{R}_4\mathrm{Ni}_3\mathrm{O}_{10}$ (R = La, Pr)--have drawn significant attention because (i) they can be synthesized in bulk, crystalline form~\cite{Crystal_growth,15,Pr4310_psc,Pr4310_metal_transition,Crystal_growth_La4310}, and (ii) they are presumed to be a multiorbital system, providing a contrast to cuprates and the infinite layer nickelates.
	Charge density wave (CDW) order has been established in these nickelate materials at ambient pressure~\cite{24,15,14,41,35,37,40,39,38,20}. High-pressure studies further reveal a global phase diagram in which increasing pressure systematically suppresses the CDW order~\cite{18,22,35,36,37,38,39,40,20}. Remarkably, pressure-induced superconductivity has been reported with critical temperatures reaching up to 80 K, exceeding the boiling point of liquid nitrogen~\cite{17,18,19,20,21,22,23,24,15}.
	These materials share key structural motifs with cuprates but differ in electronic configuration, multiorbital character, and competing ordered states, providing a distinct platform for probing pairing mechanisms in unconventional superconductivity.
	
	Despite rapid progress in the study of pressure-induced superconductivity in nickelates, two fundamental questions remain unresolved. First, the bulk nature of the superconductivity remains under debate. Although zero resistance has been observed in many samples, a full shielding effect has rarely been reported, except in cases where helium gas is used as the pressure-transmitting medium~\cite{21,15}. Other studies instead report very low magnetic shielding fractions~\cite{14,10}, raising the possibility that superconductivity may occur only in interfacial or filamentary regions. Second, once the bulk nature of pressure-induced superconductivity is firmly established, it becomes essential to elucidate the pairing mechanism--for example, by determining the gap symmetry and amplitude--particularly in light of the close structural and electronic parallels between nickelates and cuprates.
	%With electron-phonon coupling under the BCS framework widely ruled out~\cite{32}, recent theoretical studies proposed an alternative scenario, where superconductivity is mediated by spin fluctuations and likely displays  extended s$_\pm$-wave pairing symmetry~\cite{33, 34}. Other models suggest alternative mechanisms such as $d$-wave pairing or local-moment-driven pairing, based on multiorbital $t-J$, or Hubbard models~\cite{47, 48, 49, 50, 51, 52, 53, 54}. 
	%This is a crucial clue for understanding the emergence of superconductivity in these materials.
	
	Ultrafast pump-probe spectroscopy has emerged as a powerful and versatile tool for addressing these open questions, providing direct access to nonequilibrium quasiparticle (QP) and condensate dynamics complementary to static probes. For example, ultrafast QP dynamics measurements have revealed key spectroscopic signatures of $d$-wave pairing symmetry and determined a superconducting gap of $2\Delta \approx 1.2~\mathrm{THz}$ ($\sim 5~\mathrm{meV}$) in infinite-layer nickelates~\cite{n1}. More recently, THz 2D spectroscopy has independently confirmed $d$-wave pairing through the observation of strongly damped Higgs collective modes, closely resembling cuprate-like nonequilibrium responses \cite{n2}.
	Moreover, a growing number of ultrafast pump–probe studies on layered nickelates under high pressure have explored the interplay between superconductivity and density-wave orders~\cite{7,8,13}. This approach enables femtosecond-resolved tracking of QP relaxation, phonon bottlenecks, and critical slowing down near broken-symmetry gap openings, offering unique sensitivity to fluctuating density-wave orders and transient superconducting states. However, none of these prior ultrafast pressure studies have simultaneously incorporated high magnetic fields~\cite{11,12}. 
	
	We highlight three key aspects that underscore the need to develop ultrafast magneto-pressure spectroscopy, which is essential for establishing the bulk nature and nonequilibrium QP physics of pressure-induced superconductivity in nickelates. Resolving this ongoing debate is a critical prerequisite for uncovering their pairing symmetry and gap structure. 
	First, magnetic-field-dependent measurements under pressure provide a powerful tuning parameter for following the evolution of competing phases. 
	%and separating superconductivity from density-wave orders through their distinct magentic field responses.
	Second, the observed critical slowing down of ultrafast quasiparticle dynamics near correlation gap openings may also arise from non-superconducting orders; magnetic-field sensitivity is therefore crucial for confirming a superconducting origin and quantifying quasiparticle participation in the superfluid response.
	Third, magneto-pump-probe studies of bulk superconductivity have began to reveal distinct hallmark such as field-dependent pre-bottleneck dynamics driven by vortex-assisted quasiparticle trapping and Cooper-pair breaking within vortex cores. 
	Such vortex-quasiparticle dynamics are essential for further establishing the robust nature of pressure-induced superconductivity in the nonequilibrium regime, complementing equilibrium probes. If the pressure-induced nickelate superconductivity is bulk and cuprate-like, it should be Type-II and exhibit vortex trapping dynamics between the lower and upper critical fields $H_{c1}$ and $H_{c2}$; the absence of such signatures therefore places strong constraints on its superconducting volume fraction and coherence.
	
	In this article, we developed ultrafast magneto-pressure spectroscopy platform capable of operating under salient conditions--simultaneously high pressure via a diamond anvil cell (DAC) up to 40~GPa, magnetic fields up to 7~T, and cryogenic temperature down to 5~K. We applied this unique ultrafast technique to study the pressure-induced superconductivity and evolution of the competing CDW order in $\mathrm{Pr}_4\mathrm{Ni}_3\mathrm{O}_{10}$--realizing, to our knowledge, the first magneto-pressure measurement of femtosecond-resolved dynamics in any material. This technical advance enables direct examination of superconducting QP kinetics in regimes inaccessible to conventional ultrafast experiments. 
	These measurements reveal how pressure and magnetic field reshape nonequilibrium quasiparticle dynamics, clarifying the competition between charge order and incipient superconducting correlations in $\mathrm{Pr}_4\mathrm{Ni}_3\mathrm{O}_{10}$. In particular, the absence of clear magnetic field dependence and vortex-related QP dynamics--contrasting sharply with our controlled bulk superconducting results--indicates that any superconductivity remains non-bulk or highly limited in volume, highlighting ultrafast magneto-pressure spectroscopy as a sensitive discriminator of emergent correlated phases.
	
	%Together, these results establish nonequilibrium QP dynamics under magnetic field as a decisive hallmark for distinguishing true bulk superconductivity from incipient or filamentary superconducting correlations under pressure.
	
	%This unique system enables direct access to nonequilibrium quasiparticle and correlation dynamics controlled by pressure and magnetic field. Our measurements on $\mathrm{Pr}_4\mathrm{Ni}3\mathrm{O}{10}$ reveal a progressive suppression of the CDW gap with increasing pressure and provide new insight into the elusive, low-volume-fraction superconductivity that emerges in this regime. These results establish a dynamic framework for understanding how CDW order and superconductivity compete and evolve under multiple tuning parameters. More broadly, they demonstrate the power and versatility of ultrafast magneto-pressure spectroscopy as a frontier approach for probing and controlling competing orders and emergent quantum phases in correlated electronic materials.
	%Your manuscript should contain all of the numbered sections specified in this template: Introduction, Results, Discussion, Materials and Methods.
	
	%The manuscript should start with a brief introduction that lays out the problem addressed by the research and describes the paper’s importance. The scientific question being investigated should be described in detail. The introduction should provide sufficient background information to make the article understandable to readers in other disciplines and provide enough context to ensure that the implications of the experimental findings are clear. 
	
	\section{Materials and Methods}
	
	High-quality single crystals have been grown, exhibiting excellent crystallinity and minimal stacking faults-essential prerequisites for probing intrinsic phase behavior.
	%\subsection{The preparation of the single crystal and the diamond anvil cell}
	Single crystal samples of $\mathrm{Pr}_4\mathrm{Ni}_3\mathrm{O}_{10}$ used in this work were grown at Argonne National Laboratory under 140 bar of O$_2$ using a SciDre GmbH  optical-image floating-zone furnace (model HKZ)% at Argonne National Laboratory
	, as detailed in Ref.~\cite{42}. A comprehensive characterization of their intertwined charge density wave and spin density wave behavior at ambient pressure is available in a recent study~\cite{43, Intertwined}. 
	
	The trilayer RP nickelate $\mathrm{Pr}_4\mathrm{Ni}_3\mathrm{O}_{10}$ offers several advantages for our study. The crystal structure of $\mathrm{Pr}_4\mathrm{Ni}_3\mathrm{O}_{10}$ at ambient pressure and temperature is illustrated in Fig.~1(b). 
	The unit cell consists of three Ni-O layers that form a trilayer structure through the sharing of apical oxygen atoms (not shown). The high-pressure phase exhibits a distinct structural arrangement, indicating substantial modifications in lattice symmetry and local bonding configurations.
	Compared with its La-based counterparts, $\mathrm{Pr}_4\mathrm{Ni}_3\mathrm{O}_{10}$ crystallizes in a well-defined monoclinic structure $(P2{1}/a)$, which transforms into a tetragonal $(I4/mmm)$ phase under high pressure, $P > 35\,\mathrm{GPa}$
	~\cite{15}. The system exhibits a pronounced CDW transition near $T_\mathrm{CDW}\sim$158~K under ambient pressure~\cite{24,15,14,35}. Under compression for the bulk sample, the CDW order is progressively suppressed, giving rise to a superconducting dome with a maximum $T_\mathrm{c}$ approaching 40~K at pressure, $P \approx 35\,\mathrm{GPa}$
	~\cite{24,15,35}. In addition, the presence of Pr 4f electrons introduces extra orbital degrees of freedom and potential 4f-3d hybridization, enriching the system’s low-energy electronic landscape~\cite{hybridization}. It is also conceivable that Pr 4f-O 2p hybridization plays a role, analogous to scenarios discussed in Pr-based cuprates such as $\mathrm{PrBa}_2\mathrm{Cu}_3\mathrm{O}_7$~\cite{Ku2002JAP,Ruiz2022NatCommun}. Such coupling, particularly involving the apical oxygen atoms in the Pr rocksalt layer, may influence the electronic structure and could potentially contribute to the stabilization of the density-wave state. While this possibility remains speculative, it highlights an additional avenue for understanding the interplay between rare-earth orbitals and correlated electronic order in this system.
	
	\begin{figure}[t]
		\centering
		\includegraphics[width=0.9\textwidth]{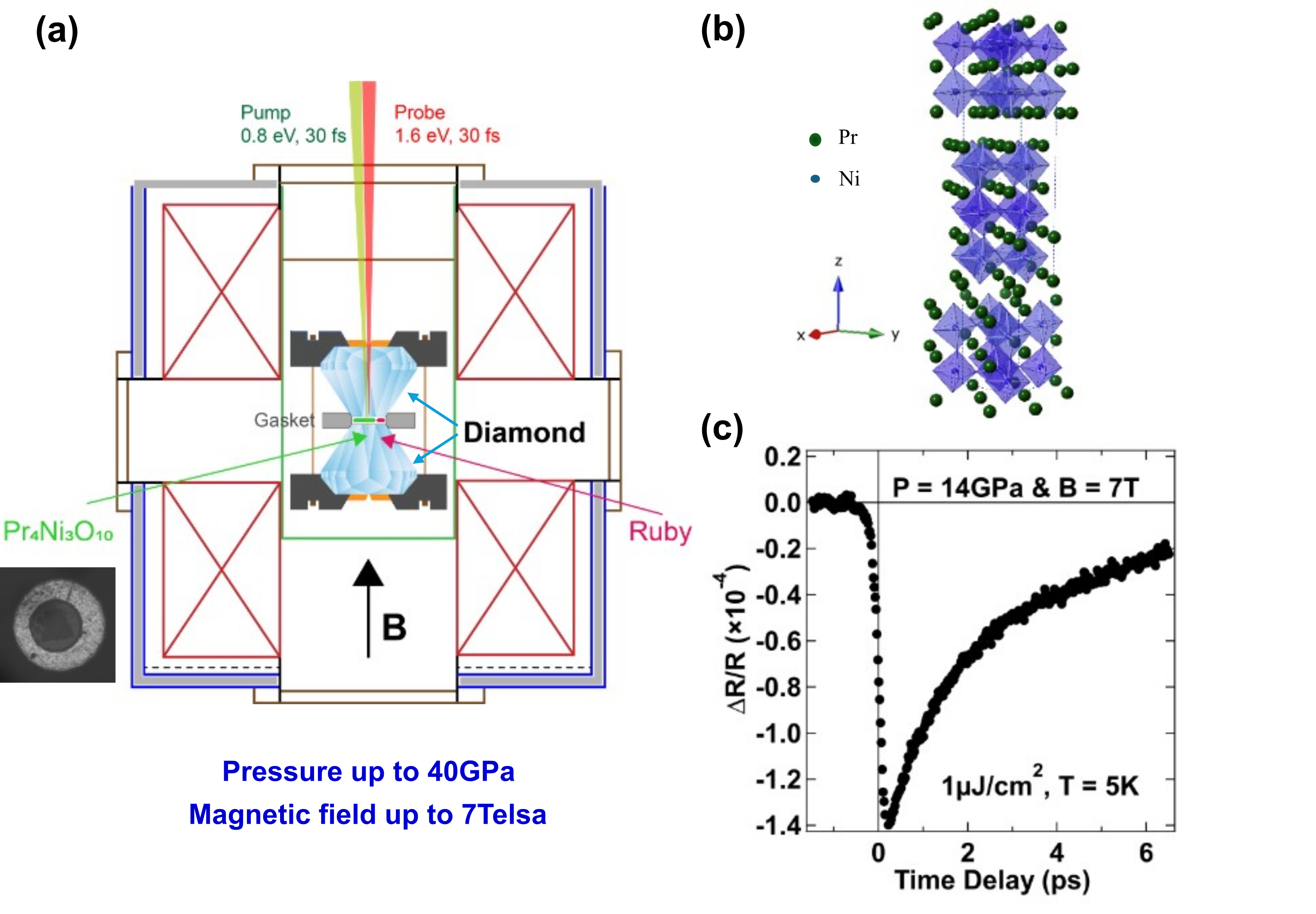}
		\caption{\textbf{Optical pump--probe measurements of Pr$_4$Ni$_3$O$_{10}$ under high pressure and magnetic field.}
			(a) Schematic of the optical pump--probe experimental setup with a diamond anvil cell (DAC) housed in a magnetic cryostat. An optical image of the sample inside the pressure cell is provided in the left. 
			(b) Crystal structure of $\mathrm{Pr}_4\mathrm{Ni}_3\mathrm{O}_{10}$ at ambient pressure, where three Ni--O layers form a trilayer structure within each unit cell. 
			(c) A representative $\Delta R/R$ signal measured at a temperature $T = 5$~K, pressure $P = 14$~GPa, magnetic field $B = 7$~T, and pump fluence of $1\,\mu\mathrm{J}/\mathrm{cm}^2$.}
		\label{fig1:main}
	\end{figure}
	
	As illustrated in Fig.~1(a), high-pressure magneto-pump-probe spectroscopy experiments were performed using a non-magnetic Be-Cu DAC ~\cite{44} equipped with standard-cut type Ia diamonds with 400 $\mu$m culets, allowing access to pressures up to 40 GPa. The sample was prepared as a thin flake, approximately 120 × 120 × 10 $\mu$m$^3$ in size (inset), with a shiny surface, and loaded into a pre-indented tungsten gasket along with small ruby spheres to act as manometers. The diameter of the sample chamber was around half the size of the culet. Nujol mineral oil was used as the pressure-transmitting medium to ensure consistency with our previous high-pressure studies on $\mathrm{Pr}_4\mathrm{Ni}_3\mathrm{O}_{10}$~\cite{14}. The pressure was determined using the position of the ruby R1 fluorescence line, following standard calibration procedures~\cite{45, 46}. The DAC was finally mounted onto the sample holder of a magneto-cryostat using a custom-made brass adapter. 
	%of the Quantum Design OptiCool system using a custom-made brass adapter.

	%\subsection{Time-resolved transient reflectivity experiments}
	Ultrafast reflectivity measurements were performed using femtosecond laser pulses with pump and probe photon energies at 0.8 eV and 1.6 eV, respectively, at temperatures from 5~K to 200~K, magnetic fields up to 7~T, and pressure up to 40~GPa (Fig.~1(a)) (see Supplementary Materials for details). The pump and probe beams were focused collinearly onto the sample in the DAC, under a perpendicular magnetic field to the sample surface. Pump-induced reflectivity changes $\Delta R/R$ were detected using a balanced detector and lock-in amplifier as discussed in~\cite{5,55}. The pump–probe time delay was controlled using a mechanical delay stage. The pump fluence was varied with neutral-density filters, while the probe fluence was kept at least an order of magnitude lower to minimize perturbation of the sample. %The schematic of our experimental setup was shown in Fig.~1(a), where the DAC is enclosed in the magnetic cryostat.

	\section{Results and Discussions}
	
	%Ultrafast dynamics has been shown to uniquely characterize superconducting condensates in the nonequilibrium regime, exhibiting hallmark phonon-bottleneck behavior at zero magnetic field and a magnetic-field-dependent pre-bottleneck regime at finite field. To probe such signatures under extreme conditions, we perform femtosecond pump–probe spectroscopy under simultaneous high pressure and magnetic field—realizing, to our knowledge, the first magneto-pressure measurement of femtosecond-resolved dynamics in any material. This technical advance enables direct examination of superconducting QP kinetics in regimes inaccessible to conventional ultrafast experiments.
	
	%Within the phonon-bottleneck regime, QP recombination generates high-energy phonons (HEPs) that subsequently re-break Cooper pairs, forming a feedback loop that suppresses net recombination. Consequently, QPs persist over extended timescales, producing a long-lived $\Delta R/R$ response with decay times on the order of hundreds of picoseconds for $s$-wave superconductors and somewhat shorter lifetimes for $d$-wave symmetry. In infinite-layer nickelates, which likely host a $d$-wave gap, recent ultrafast measurements report QP lifetimes of approximately $100~\mathrm{ps}$ in the superconducting state~\cite{n1}. Under an applied magnetic field, pre-bottleneck QP dynamics emerge from vortex states in Type-II superconductors, giving rise to a distinctive field-dependent buildup of the $\Delta R/R$ response, with slow rise times on the order of 10 picoseconds.

	Figure~1(c) presents representative $\Delta R/R$ temporal dynamics of $\mathrm{Pr}_4\mathrm{Ni}3\mathrm{O}{10}$ measured at $T = 5~\mathrm{K}$, $P = 14~\mathrm{GPa}$, and $B = 7~\mathrm{T}$. The response exhibits a femtosecond-scale buildup governed by the pump pulse duration and rapid electronic thermalization, followed by a decay within $\lesssim 10$~ps. According to zero-field cooling (ZFC) DC magnetic susceptibility measurements from our previous work~\cite{14}, $\mathrm{Pr}_4\mathrm{Ni}_3\mathrm{O}_{10}$ is expected to exhibit some degree of superconductivity at $P = 14$~GPa. If superconducting quasiparticle kinetics were present under these conditions, one would anticipate a long-lived $\Delta R/R$ signal characterized by field-dependent rise times of $\sim 10$~ps and slow decay times approaching $\sim 100$~ps, because under an applied magnetic field, pre-bottleneck QP dynamics emerge from vortex states in Type-II superconductors, giving rise to a distinctive field-dependent buildup of the $\Delta R/R$ response, with slow rise times on the order of 10 picoseconds. However, no long-lived component or field-dependent rise dynamics are observed. Therefore, the measured $\Delta R/R$ behavior is inconsistent with superconducting quasiparticle bottleneck kinetics under the present experimental conditions and instead points toward CDW dynamics, as discussed below.
	
	Figure~2 demonstrates pronounced critical slowing down of QP relaxation at the CDW critical temperature, which collapses under applied high pressure.  
	%Ultrafast pump-probe spectroscopy is a unique and nondestructive technique for investigating systems driven out of equilibrium. 
	The ultrafast dynamics within the CDW phase can be described by a phonon-bottleneck mechanism analogous to that in superconducting condensates~\cite{1,3,4}. 
	%In this picture, nonequilibrium phonons impede the recombination of photoexcited carriers across the CDW gap. 
	Accordingly, the Rothwarf–Taylor (RT) model~\cite{2}, originally developed for superconducting kinetics, can be extended to describe ultrafast quasiparticle dynamics in the CDW phase.
	%The pump pulse excites the system, driving it out of equilibrium, while the probe pulse monitors the relaxation dynamics at delayed times. 
	Upon photoexcitation, the pump injects hot carriers that rapidly relax through electron-electron and electron–phonon scattering, generating nonequilibrium QPs and high-frequency phonons (HFPs). In CDW systems, the condensate consists of bound QP pairs (excitons), which can be dissociated by enhanced Coulomb screening from hot carriers or by absorption of HFPs with energies exceeding the CDW gap. These processes drive the initial depletion of the condensate, followed by a recovery governed by the coupled dynamics of QPs and HFPs. Importantly, as the system approaches the CDW transition temperature, the gap softens and the lifetime of HFPs increases, reducing the efficiency of QP recombination into the ground state and leading to a pronounced slowing of the relaxation dynamics. This critical slowing down reflects the diverging relaxation timescales associated with order-parameter fluctuations near the CDW phase transition and serves as a hallmark of the onset of macroscopic CDW order~\cite{6}.
	
	\begin{figure}[h]
		\centering
		\includegraphics[width=1.0\textwidth]{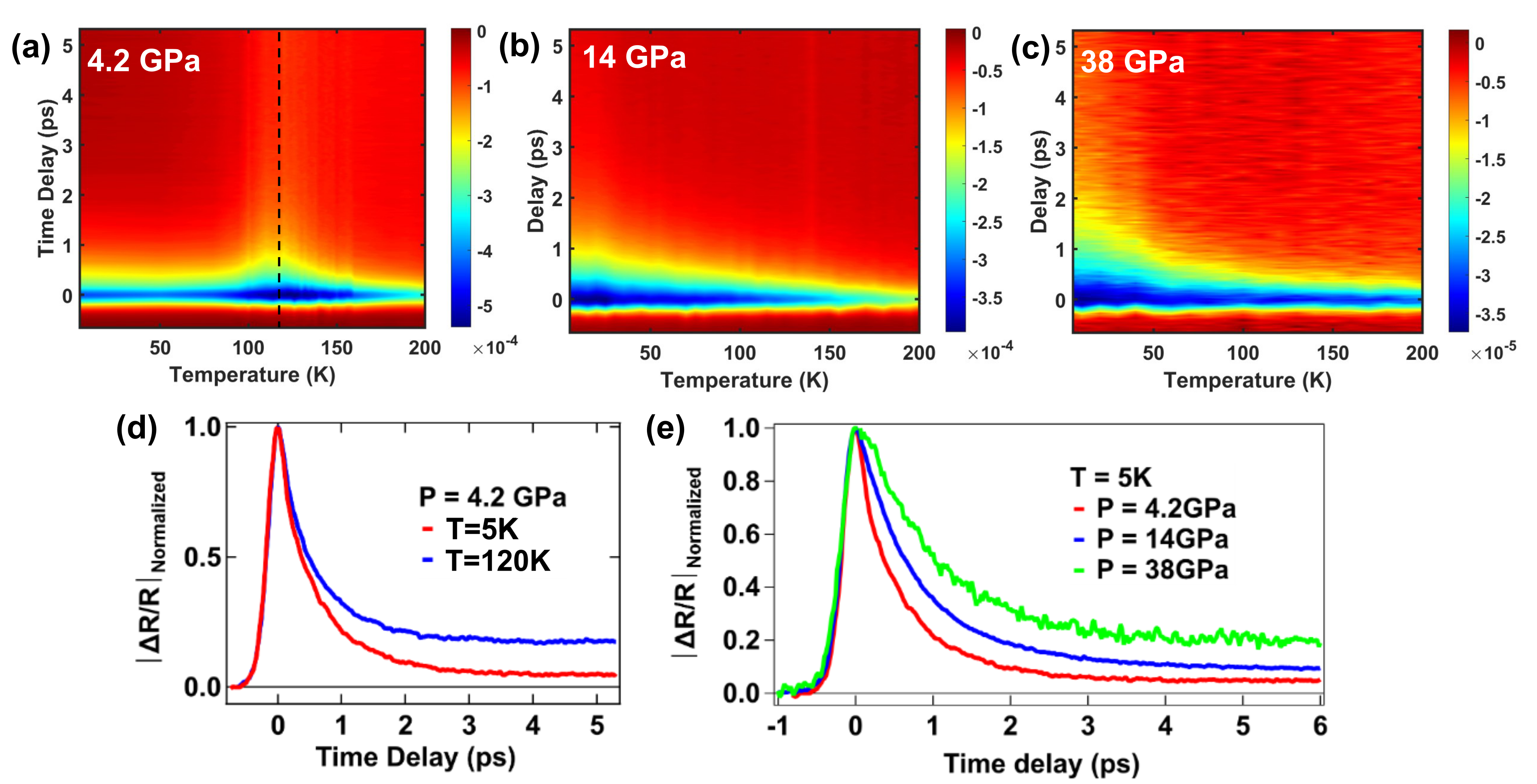}
		\caption{\textbf{Pressure-dependent $\Delta R/R$ in Pr$_4$Ni$_3$O$_{10}$.}
			(a)--(c) 2D false-color plots of temperature-dependent $\Delta R/R$ at 4.2, 14, and 38~GPa, respectively. 
			(d) Normalized $\Delta R/R$ at 4.2~GPa measured at 5~K and 120~K, highlighting distinct dynamics at different temperatures. 
			(e) Normalized $\Delta R/R$ at $T = 5$~K for 4.2, 14, and 38~GPa, showing the slowing of relaxation dynamics at higher pressures. 
			All data were measured at a pump fluence of $7\,\mu\mathrm{J}/\mathrm{cm}^2$.}
		\label{fig2:main}
	\end{figure}
	
	Figure~2(a) shows a two-dimensional false-color plot of the temperature-dependent $\Delta R/R$ signals measured at 4.2~GPa. A pronounced enhancement and slowing down of the photoinduced reflectivity response is observed near $\sim 120$~K. This temperature is lower than the equilibrium CDW transition temperature ($\approx 140$~K at 4.2~GPa) reported in Ref.~[1]. We attribute this $\sim 20$~K downward shift primarily to pump-induced electronic and lattice heating at the employed fluence of 7 $\mu\mathrm{J/cm}^2$. At this excitation level, the effective temperature of the optically probed volume is expected to exceed the nominal cryostat setpoint, thereby renormalizing the apparent transition temperature.
	
	In contrast, at 14~GPa (Fig.~2(b)), no comparable anomaly is resolved. Although Ref.~[1] reports an equilibrium transition temperature $\approx 110$~K at this pressure, the response remains relatively featureless over the measured temperature range. Given the heating effects established at 4.2~GPa, it is plausible that pump-induced temperature elevation at 14~GPa smears or obscures the critical dynamics, effectively shifting the system across the transition during measurement.
	
	Moreover, at 38~GPa (Fig.~2(c)), the temperature dependence is entirely featureless across the full range studied. In this case, the absence of an anomaly is unlikely to arise from heating alone and instead is consistent with the reported suppression of the CDW phase at high pressure in favor of an incipient superconducting state~[1]. This systematic evolution highlights pressure as an effective tuning parameter for weakening competing CDW order and potentially stabilizing superconductivity in these correlated nickelates.
	
	To highlight the prolonged relaxation near the CDW transition temperature at 4.2~GPa, we compare the normalized $\Delta R/R$ at $T = 5$~K and 120~K, as shown in Fig.~2(d). At 120~K, the critical slowing of the relaxation manifests as a clearly prolonged decay.
	Figure~2(e) shows $\Delta R/R$ measured at $T = 5~\mathrm{K}$ for pressures of 4.2, 14, and 38~GPa. A clear slowing of the decay dynamics with increasing pressure is observed. At 4.2~GPa, this slowing can be attributed to the CDW gap, while at higher pressures—where the CDW may be partially or fully suppressed—the extended relaxation likely reflects modifications of QP dynamics independent of long-range CDW order.
	%This trend further suggests that the system evolves toward a competing ordered state such as superconductivity at high pressure.
	The pronounced increase in quasiparticle lifetimes near 120 K at lower pressure, along with their suppression at higher pressures, hints at the emergence of a potentially new electronic phase distinct from the suppressed CDW. Such behavior is reminiscent of precursor phenomena observed near superconducting transitions, where enhanced electronic correlations lead to superconducting gap formation and slowed relaxation dynamics due to phonon-bottleneck~\cite{5,NB}. Therefore, the progressive slowing of relaxation with increasing pressure should reflect the development of an alternative correlated phase, driven by changes in the electronic structure and interactions under lattice compression. This finding highlights the salient role of pressure in tuning the balance between competing orders in $\mathrm{Pr}_4\mathrm{Ni}_3\mathrm{O}_{10}$.
	
	To reduce electronic and lattice heating effects as observed in Fig. 2 under $7~\mu\mathrm{J/cm}^2$ fluence, the measurements in Figs.~3(a)-3(b) were performed at a much lower fluence of $1~\mu\mathrm{J/cm}^2$, enabling reliable investigation of the pressure-induced phase with minimal heating. By comparing Figs. 3(a) and 3(b), it is evident that the relaxation becomes increasingly prolonged at lower temperature and/or higher pressure, despite the reduced $\Delta R/R$ amplitude at higher pressure. Such pressure-induced slowing down is commonly interpreted as signatures of a tendency toward a new ordered phase that is not fully realized, characterized by an incipient opening of a correlation gap~\cite{13}. To better illustrate this trend, we compare the temporal traces at 0, 14, and 38~\(\mathrm{GPa}\), plotted on a logarithmic scale at 5 K, as shown in Fig.~3(c). These results corroborate a tendency toward the development of a new electronic state following the suppression of the CDW phase.
	To gain quantitative insight into the temperature-dependent relaxation dynamics, we fit the $\Delta R/R$ for the 14~\(\mathrm{GPa}\) data (Fig. 3(a)), which can be reproduced well by a single-exponential decay function. The extracted amplitude $A$ and relaxation time $\tau$ are summarized in Figs.~3(d) and 3(e), respectively. Note that a similar trend is also observed for 38 GPa data (not shown). The amplitude $A$ increases slightly with temperature up to $\sim 40~\mathrm{K}$, remains nearly constant up to 100 K, and is strongly suppressed above $\sim 100~\mathrm{K}$, while the relaxation time $\tau$ decreases markedly with increasing temperature, as expected. 
	Note that a single-exponential fit was used for Fig.~3 ($1~\mu\mathrm{J}$, $14~\mathrm{GPa}$), whereas a bi-exponential function was required for Fig.~4 ($0.3~\mu\mathrm{J}$, $0~\mathrm{GPa}$) to be discussed. At high pressure the relaxation becomes significantly prolonged, and within the experimentally focused time window the decay is dominated by a single slow channel, making a single-exponential description sufficient. In contrast, at ambient pressure the transient clearly exhibits both fast and slow relaxation components, necessitating a bi-exponential fit.

	%While double-exponential models can be used when two distinct relaxation timescales are resolved, the temporal window of the present measurement (6 ps) does not extend sufficiently to capture slower dynamics.
	
	\begin{figure}[h]
		\centering
		\includegraphics[width=1.0\textwidth]{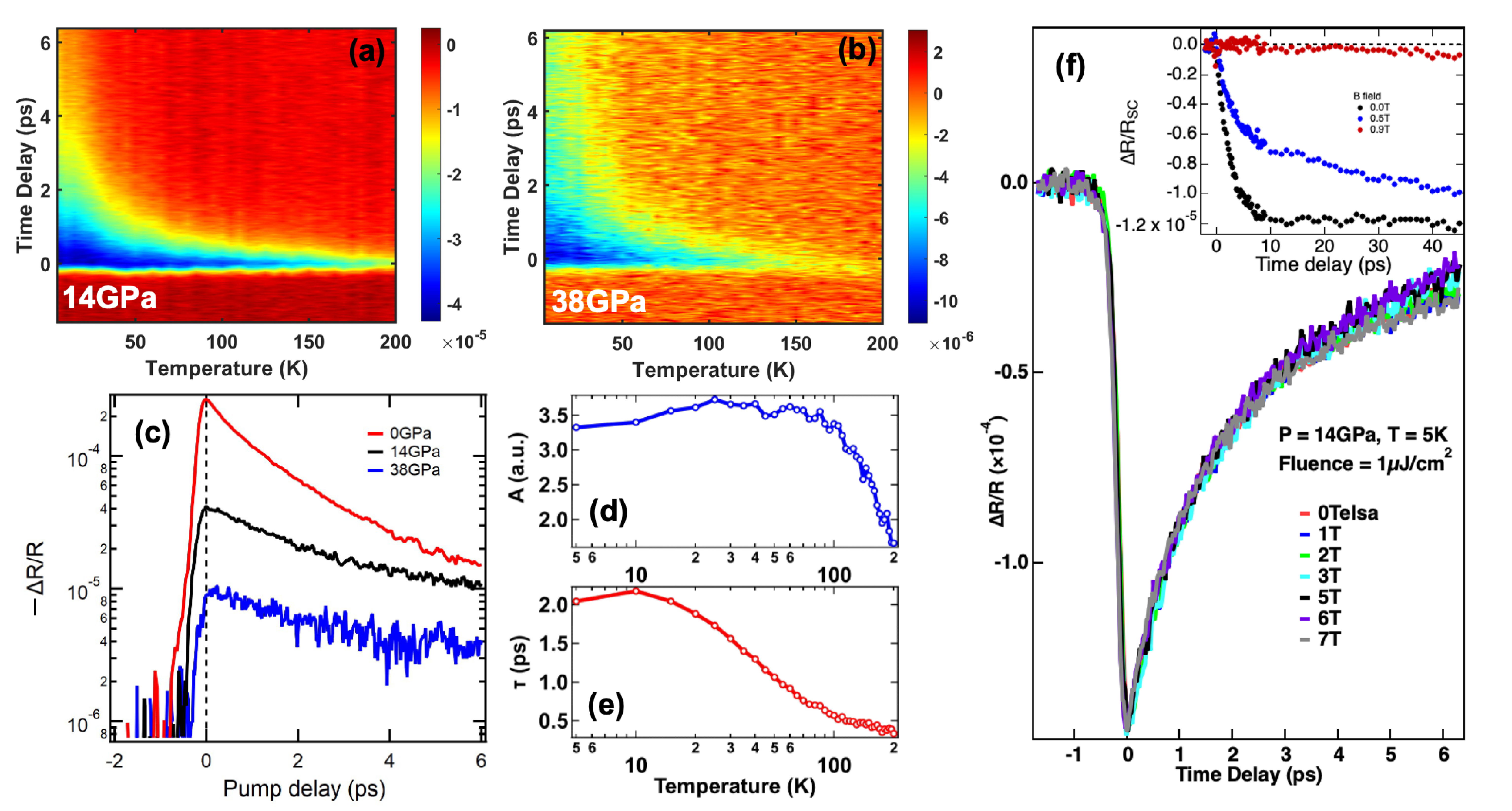}
		\caption{\textbf{Pressure- and magnetic-field dependent $\Delta R/R$ in Pr$_4$Ni$_3$O$_{10}$.}
			(a)--(b) 2D false-color plots of temperature-dependent $\Delta R/R$ at (a) 14~GPa and (b) 38~GPa with a pump fluence of $1\,\mu\mathrm{J}/\mathrm{cm}^2$. 
			(c) $\Delta R/R$ measured at $T = 5$~K for 0, 14, and 38~GPa with a fluence of $1\,\mu\mathrm{J}/\mathrm{cm}^2$. 
			Temperature dependence of fitted (d) amplitude $A$ and (e) relaxation time $\tau$ are obtained using a single-exponential decay model for the data in (a). 
			(f) $\Delta R/R$ at selected magnetic fields, illustrating weak magnetic-field dependence of the photoinduced dynamics below 7 T. 
			The inset shows $\Delta R / R_{\mathrm{SC}}$ for niobium sample measured at $T \sim 2.3$~K, $1\,\mu\mathrm{J}/\mathrm{cm}^2$ fluence, and at selected magnetic fields: $B$=0, 0.5, and 0.9 T, where vortex states are present with finite fields.}
		\label{fig3:main}
	\end{figure}
	
	To achieve a direct assessment of the presence or absence--and the nature--of a possible superconducting state, we measure magnetic-field-dependent $\Delta R/R$ at 14~\(\mathrm{GPa}\), 5~\(\mathrm{K}\), and \(1~\mu\mathrm{J/cm}^2\), as shown in Fig.~3(f). No discernible changes in $\Delta R/R$ signal are observed for magnetic fields up to 7~T. The weak field dependence indicates that the superconducting-like signatures we observe in Figs. 2 and 3 do not arise from bulk superconductivity. In such a state, the application of an external magnetic field is expected to weaken superconductivity and produce discernible changes in the transient optical response. 
	A defining hallmark of nonequilibrium dynamics in Type-II superconductors is the magnetic-field-dependent, delayed buildup of the $\Delta R/R$ signal, reflecting the crucial role of vortex-controlled QP trapping process. By contrast, the absence of any comparable magnetic-field-dependent buildup or suppression of $\mathrm{Pr}_4\mathrm{Ni}_3\mathrm{O}_{10}$ in our data indicates that no clear superconducting signatures are observed below 7 T in the present pump--probe measurements. As a comparison, such behavior is clearly observed in bulk superconducting states of a thick Niobium film below its critical temperature~\cite{NB} (inset of Fig.~3(f)), where the transient response at 0~T differs markedly from that at 0.5~T, and upon further increasing the magnetic field to 0.9~T, the superconducting signal is significantly suppressed (red trace), consistent with magnetic-field-induced destruction of superconductivity. Taken together with superconductivity established by temperature-dependent ZFC DC magnetic susceptibility ~\cite{14}, this disparity suggests that the superconducting response may not be bulk in nature.
	%, in which superconducting regions are confined to isolated or weakly connected domains lacking global phase coherence. 
	These observations highlight a subtle interplay among pressure, temperature, and superconducting correlations in the nickelate system, suggesting that superconductivity, if present, emerges in an inhomogeneous form lacking global phase coherence and remains largely insensitive to magnetic fields up to 7~T. This conclusion is further supported by temperature-dependent ZFC DC magnetic susceptibility measurements in Ref.~\cite{14}, which display features consistent with the onset of superconducting correlations. 
	%The same ZFC DC magnetic susceptibility measurements~\cite{14} estimate the superconducting volume fraction to be on the order of 10\%, indicating that the superconducting state--if present--is likely inhomogeneous or filamentary rather than bulk. 
	%However, the exact microscopic origin of the observed decay behavior remains to be clarified.
	%This behavior is consistent with the gradual suppression of superconducting regions in a filamentary superconductor, where thermal fluctuations weaken the coherence or population of quasiparticles associated with superconducting filaments. As the temperature increases, the contribution of these regions to the overall transient reflectivity signal diminishes, resulting in a reduced amplitude. 
	%This trend suggests that the relaxation process becomes more efficient at elevated temperatures, potentially due to enhanced scattering or energy dissipation mechanisms. The temperature dependence is consistent with a relaxation pathway influenced by an intrinsic energy scale. While the bulk of the sample remains non-superconducting, the dynamics appear to be partially governed by the localized superconducting phase. 

	To investigate the pump-fluence dependence of the $\Delta R/R$ dynamics, Figs.~4(a)–4(c) present 2D false-color maps of the temperature-dependent $\Delta R/R$ at ambient pressure for pump fluences of 7, 1, and 0.3$~\mu\mathrm{J/cm}^2$, respectively. A pronounced fluence dependence of the CDW transition temperature is observed: in the low-excitation regime ($0.3$ and $1~\mu\mathrm{J/cm}^2$), $T_{\mathrm{CDW}}$ remains close to its equilibrium value of $\sim$158 K, whereas at the highest fluence applied ($7~\mu\mathrm{J/cm}^2$), it is substantially suppressed to approximately 120~K.
	%(approximately 155~K) 
	The suppression of \(T_{\mathrm{CDW}}\) with increasing excitation fluence likely originates from two complementary mechanisms. First, photoexcited carriers transiently screen the periodic lattice distortion that stabilizes the CDW state, thereby weakening the electron--phonon coupling and destabilizing the charge order. Second, electronic heating at higher fluence can locally elevate the electronic temperature above the equilibrium transition temperature, leading to partial melting of the CDW phase. Prior ultrafast spectroscopy studies of superconductors using similar laser conditions have established that local lattice heating contributes significantly to the signal only at fluences above 10$~\mu\mathrm{J/cm}^2$ \cite{2,3}.    
	Together, these effects account for the pronounced reduction of \(T_{\mathrm{CDW}}\) under strong excitation. In addition, ultrafast pump-probe measurements are not sensitive to slow, bulk thermal anchoring. Therefore, these observations confirm that the low--pump--fluence regime below \(1~\mu\mathrm{J/cm}^2\), as employed in Fig.~3, is suitable for probing pressure--induced superconductivity while minimizing perturbations to the underlying electronic order.
	
	\begin{figure}[h]
		\centering
		\includegraphics[width=0.9\textwidth]{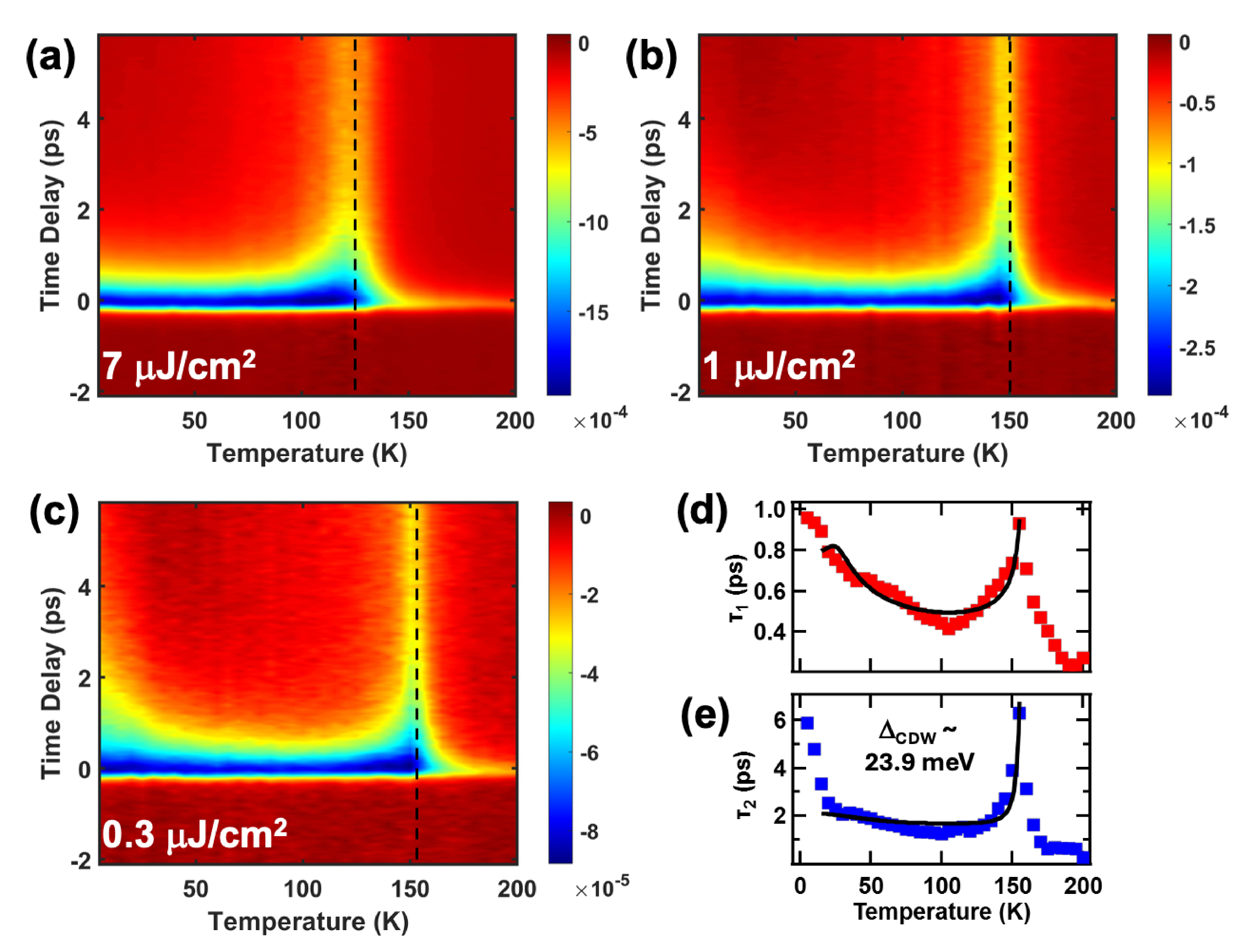}
		\caption{\textbf{Fluence-dependent relaxation dynamics and CDW gap extraction at ambient pressure.}
			(a)--(c) 2D false-color plots of the temperature-dependent $\Delta R/R$ measured at 0~GPa for excitation fluences of $7\,\mu\mathrm{J}/\mathrm{cm}^2$, $1\,\mu\mathrm{J}/\mathrm{cm}^2$, and $0.3\,\mu\mathrm{J}/\mathrm{cm}^2$, respectively. 
			(d)--(e) Corresponding relaxation times ($\tau_1$ and $\tau_2$) extracted by fitting the data in (c) using a bi-exponential decay model. Both $\tau_1$ and $\tau_2$ are fitted with Eq.~\eqref{eq:tau}, yielding a CDW gap $\Delta_{\mathrm{CDW}}$ of approximately 23.9~meV.}
		\label{fig4:main}
	\end{figure}

	To quantify the relaxation dynamics, the data measured at the smallest pump fluence (Fig.~4(c)) were fitted. However, unlike the behavior shown in Fig.~3, the data in Fig. 4(c) cannot be adequately described by a single-exponential decay. Instead, a bi-exponential decay model is required to reproduce the results satisfactorily. The extracted fast and slow relaxation time constants, \(\tau_1\) and \(\tau_2\), are summarized in Figs.~4(d) and 4(e), respectively. Both \(\tau_1\) and \(\tau_2\) exhibit pronounced divergences upon approaching \(T_{\mathrm{CDW}}\), consistent with the critical slowing down expected near a continuous phase transition, as discussed above.  In the weak-perturbation regime, where the density of photoexcited quasiparticles is small, the relaxation dynamics can be well described within a BCS-like, correlation gap framework. Assuming a mean-field temperature dependence of the CDW gap,
	\begin{equation}
		\Delta(T) = \Delta_{\mathrm{CDW}} \sqrt{1 - \frac{T}{T_{\mathrm{CDW}}}},
		\label{eq:gap}
	\end{equation}
	the temperature dependence of the relaxation time can be expressed as
	\begin{equation}
		\tau(T) \propto \frac{\ln\!\left(g + e^{-\Delta(T)/k_{\mathrm{B}} T}\right)}{\Delta(T)^2},
		\label{eq:tau}
	\end{equation}
	where \(k_{\mathrm{B}}\) is the Boltzmann constant and \(g\) is a phenomenological parameter accounting for residual recombination channels. Simultaneous fits (black lines, Figs.~4(d)-4(e)) to the experimental \(\tau_1(T)\) and \(\tau_2(T)\) data using Eqs.~\eqref{eq:gap} and~\eqref{eq:tau} yield a CDW gap magnitude of \(\Delta_{\mathrm{CDW}} \approx 23.9\)~meV at the pump fluence of \(0.3~\mu\mathrm{J/cm}^2\) for both relaxation components. For comparison, other experimental studies on the trilayer nickelate La$_4$Ni$_3$O$_{10}$ report a CDW gap of $\Delta_{\mathrm{CDW}} \approx 50$~meV, which is somewhat larger than the value extracted from our fitting model for bulk Pr$_4$Ni$_3$O$_{10}$~\cite{13,PhysRevB.111.075140,2p56-xl41}. 
	%Note that one distinct difference is that the bi-exponential decay profile (Fig.~4(a)-4(c)) in the CDW phase is replaced by a single exponential decay profile (Fig.~3(a)-3(c))      

	%These results further support the interpretation that high excitation fluence not only suppresses \(T_{\mathrm{CDW}}\) but also reduces the magnitude of the CDW gap, consistent with a weakening of the underlying charge order.
	
	%The section should describe the experiments performed and the findings observed. The results section should be divided into subsections to delineate different experimental themes. Subheadings should either be all phrases or all complete sentences. All data must be shown either in the main text or in the Supplementary Materials. 
	
	%\section{Discussion} (JW)
	% Comments
	% Previous pump-probe studies on nickelate  \cite{7, 8, 13}\\
	% Bulk Pr4310 \cite{15}\\
	% Phonon bottleneck for CDW \cite{6}\\
	% Shuyuan's transport measurement \cite{14}
	% RT model for CDW transition.
	% The reason for using bi-exponential decay?
	% The CDW gap 27 meV measured with (0.3\,\mu\text{J}/\text{cm}^2\) is close to other papers?

	%Include a Discussion that summarizes (but does not merely repeat) your conclusions and elaborates on their implications. There should be a paragraph outlining the limitations of your results and interpretation, as well as a discussion of the steps that need to be taken for the findings to be applied. Please avoid claims of priority. 
	
	\section{Conclusion}
	Ultrafast magneto-pressure spectroscopy reveals how competing CDW and potential incipient superconducting correlations evolve in $\mathrm{Pr}_4\mathrm{Ni}_3\mathrm{O}_{10}$ under high pressure conditions. Low-fluence measurements establish a well-defined CDW transition, while critical slowing down and its pressure-induced collapse directly demonstrate the fragility of CDW order to pressure-induced lattice compression. At higher pressures, the prolonged quasiparticle lifetimes observed at low temperatures signal the suppression of CDW order and may signal the emergence of incipient superconducting correlations. However, the absence of magnetic-field-dependent pre-bottleneck dynamics rules out bulk superconductivity, pointing instead to filamentary or inhomogeneous superconducting states. More broadly, our work establishes ultrafast magneto-pressure spectroscopy as a decisive test for bulk superconductivity: a genuine bulk superconducting state must produce a measurable response in this nonequilibrium probe under magnetic field and pressure, whereas incipient or competing ordered states do not.
	
	%The ultrafast dynamic behaviors of photoexcited quasiparticles in CDW and the superconducting phase of tri-layered Pr$_4$Ni$_3$O$_{10}$ are studied with pump-induced reflectivity change under high pressure and high magnetic field. The CDW temperature of 155 K measured with low pump fluence \(0.3 \, \mu\text{J}/\text{cm}^2\) agrees well with transport measurement and low excitation energy is needed to lower the temperature induced by heating and reduce electronic screening to get CDW transition temperature. The CDW gap 23.9~meV is obtained from the fitting of temperature dependent decay time. The low temperature dynamic behaviors under magnetic field indicates that the superconducting states are not originated not from bulk but from surface or filamentary.
	
	\section*{Acknowledgments and Funding}
	The ultrafast magneto-optical experiments were supported by the National Science Foundation under Award No. 2530947.
	The high-pressure setup was supported by the U.S. Department of Energy, Office of Basic Energy Sciences, Division of Materials Sciences and Engineering. Ames National Laboratory is operated for the U.S. Department of Energy by Iowa State University under Contract No. DE-AC02-07CH11358. SH was supported, in part, by the Ames National Laboratory's Laboratory Directed Research and Development (LDRD) program. Single-crystal growth and characterization of the nickelate samples in the Materials Science Division of Argonne National Laboratory were supported by the U.S. Department of Energy, Office of Science, Basic Energy Sciences, Materials Sciences and Engineering Division.
	
	%\subsection*{General} 
	%Thank others for any contributions, whether it be direct technical help or indirect assistance. 
	
	\subsection*{Author Contributions} 
	\begin{itemize}
		\item \textbf{Jigang Wang, S. L. Bud’ko}: Supervision
		\item \textbf{Jigang Wang, S. L. Bud’ko, P. C. Canfield, Shuyuan Huyan}: Project Conceptualization
		\item \textbf{Joong-Mok Park, Zhi Xiang Chong, Shuyuan Huyan}: Data curation, Analysis, Visualization 
		\item \textbf{Shuyuan Huyan, S. L. Bud’ko, P. C. Canfield}: Preparing Pressure cell
		\item \textbf{Xinglong Chen, Daniel P. Phelan, J. F. Mitchell} Single crystal growth, characterization 
		\item \textbf{Joong-Mok Park, Zhi Xiang Chong, Shuyuan Huyan, Liang Luo, Ilias E. Perakis}: Writing – original draft
		\item \textbf{Jigang Wang}: Writing – revised draft
		\item \textbf{All authors}: Discussion, Analysis, Review \& Editing
	\end{itemize}

	\subsection*{Competing interests}
	The authors declare no competing interests.
	
	\subsection*{Data Availability}
	The data will be deposited in an online repository.
	
	%Describe any supplementary materials submitted with the manuscript (e.g., audio files, video clips or datasets). 
	
	%Please group supplementary materials in the following order: materials and methods, figures, tables, and other files (such as movies, data, interactive images, or database files). 
	
	%\medskip Example:
	%Fig. S1. Title of the first supplementary figure.
	
	%Fig. S2. Title of the second supplementary figure.
	
	%Table S1. Title of the first supplementary table.
	
	%Data file S1. Title of the first supplementary data file.
	
	%Movie S1. Title of the first supplementary movie.
	
	%\medskip
	%Be sure to submit all supplementary materials with the manuscript and remember to reference the supplementary materials at appropriate points within the manuscript. We recommend citing specific items, rather than referring to the supplementary materials in general, for example: ``See Figures S1-S10 in the Supplementary Material for comprehensive image analysis.''
	
	%A link to access the supplementary materials will be provided in the published article.
	
	%Supplementary Materials may include additional author notes—for example, a list of group authors.

	\printbibliography
	
\end{document}